\documentclass[reprint, superscriptaddress, amssymb, aps, pra, a4, showkeys]{revtex4-2}
\usepackage{graphicx}
\usepackage{hyperref}
\usepackage{natbib}
\usepackage{amsmath}
\usepackage{tabularx}
\usepackage{xcolor}
\usepackage{braket}

\begin{document}

\title{LFSR based RNG on low cost FPGA for QKD applications}

\author{Pooja Chandravanshi} \email{pooja@prl.res.in}
\affiliation{Quantum Science and Technology Laboratory, Physical Research Laboratory, Ahmedabad, India 380009.}

\author{Jaya Krishna Meka}
\affiliation{Quantum Science and Technology Laboratory, Physical Research Laboratory, Ahmedabad, India 380009.}

\author{Vardaan Mongia} \email{vardaan@prl.res.in}
\affiliation{Quantum Science and Technology Laboratory, Physical Research Laboratory, Ahmedabad, India 380009.}
\affiliation{Department of Physics, Indian Institute of Technology, Gandhinagar, India 382355.}

\author{Ravindra P. Singh} 
\affiliation{Quantum Science and Technology Laboratory, Physical Research Laboratory, Ahmedabad, India 380009.}

\author{Shashi Prabhakar}\email{shaship@prl.res.in}
\affiliation{Quantum Science and Technology Laboratory, Physical Research Laboratory, Ahmedabad, India 380009.}

\date{\today}

\begin{abstract}
Linear-feedback shift register (LFSR) based pseudo-random number generator (PRNG) has applications in a plethora of fields. The issue of being linear is generally circumvented by introducing non-linearities as per the required applications, with some being adhoc but fulfilling the purpose while others with a theoretical proof. The goal of this study is to develop a sufficiently ``random" resource for Quantum Key Distribution (QKD) applications with a low computational cost. However, as a byproduct, we have also studied the effect of introducing minimum non-linearity with experimental verification. Starting from the numerical implementation to generate a random sequence, we have implemented a XOR of two LFSR sequences on a low-cost FPGA evaluation board with one of the direct use cases in QKD protocols. Such rigorously tested random numbers could also be used like artificial neural networks or testing of circuits for integrated chips and directly for encryption for wireless technologies.
\end{abstract}


\maketitle

\section{Introduction} \label{sec:Introduction}
In any digital encryption scheme, random numbers are an indispensable resource, as proposed by Claude Shannon \cite{shannon1949communication}. The applications of random numbers are stratified from cryptography to simulations and the gambling industry. For cryptographic applications, in an ideal world, one would like to use quantum random number generators (QRNGs), which are innately random. However, pertaining to the resource intensiveness, the real-time use case for QRNGs is a little far. The next best alternatives available are pseudo random number generators (PRNGs) and chaotic random number generators. Chaotic ones cannot be used for security applications primarily because Eve can exploit the fact that there is no good way of characterising noise. Amongst PRNGs, one of the earliest, extremely fast and easy to generate known techniques for generating random numbers is using linear feedback shift registers (LFSRs). Being hardware friendly, they are used in testing for power consumption in integrated circuits. In particular, it is typically used to create test patterns for built-in self-test \cite{devika2017design}. On the low power front, LFSR based Random Number generators (LRNGs) are widely used in providing encryption to wireless technology such as Bluetooth, where they are used to encrypt frequency hopping spread spectrum systems to protect against signal jamming kind of attacks \cite{ebrahimzadeh2013frequency}.

Such random number generators are easily implementable on hardware like Field Programmable Gate Array (FPGA) \cite{madhupavani2017design} and, thus, could provide an easy access to ``randomness'' resources for ongoing applications like Quantum Key Distribution (QKD). Succinctly, QKD is the study of securely distributing keys between two communicating parties (say, Alice and Bob) using properties of photons (quanta of light). The field is proposed to takeover the current encryption technique (particularly, ubiquitously used Rivest-Shamir-Adleman (RSA) algorithm) for the latter being inefficient against upcoming quantum computers. The premise being that laws of quantum physics are more fundamental compared to increasingly accessible computational resources (built on fundamentals of physics) against an eavesdropper \cite{krishna2001test}. The major application for LRNG in QKD applications comes into the picture for prepare-and-measure (P\&M) protocols. Pertaining to the particular property of photons being used, there are different P\&M based QKD protocols such as those proposed by Bennet and Brassard in 1984 (BB84) and coherent one-way protocol. In the context of BB84 protocol, Alice selects the polarization degree of freedom as her encoding mechanism. To achieve an unbiased distribution of signals across the four polarization states, Alice requires a random bit stream based on LFSR. It generates random numbers that Alice utilizes for preparing the four polarization states in a randomized manner, ensuring equal probability and impartiality among the different polarization states. This requirement is a pre-requisite for unconditional security of any P\&M based QKD protocol. In addition to being used at Alice's end, Bob could also use LRNG sequence to choose the measurement basis by feeding this random bit stream into a liquid crystal polarisation rotator. This requirement at Bob's end is true for every QKD protocol (not restraining to P\&M protocols). This, however, is not a strict requirement as there are other alternatives with certain trade-offs. Hence, we focus on a stronger requirement of ``randomness'' resource at Alice's end.

In this work, we aim to obtain a LRNG bit stream, which is implementable on low cost, low powered FPGA boards suitable for BB84 protocol. It is varied for different parameters, say input seed value and post-processed so that it is sufficiently random for QKD applications. The paper is divided into four further sections. Section \ref{sec:Methodology} discusses preliminaries such as why such random numbers are a good candidate for QKD applications, how it works, and what classifies as a sufficiently random bit stream. In section \ref{sec:result}, we first discuss our numerical approach followed by hardware implementation of LFSR based random numbers, where we elaborate on how the randomness gets affected by introducing non-linearities. Towards the end, we show one practical application where we have used this in practical BB84 setup for preparing polarisation states randomly. In Section \ref{sec:outlook}, we list possible improvements that are hardware and application dependent. In section \ref{sec:conclusion}, we conclude the article by summarizing this study.
 
\section{Methodology} \label{sec:Methodology}
Here, we focus on generating pseudo random numbers that are sufficiently random and easily implementable on FPGA. LFSR fits this overlap naturally compared to any other pseudo random number algorithm.

\subsection{Working of LFSR} \label{sec:WorkingLFSR}
\begin{figure}[h]
    \centering
    \includegraphics[width=8.5cm]{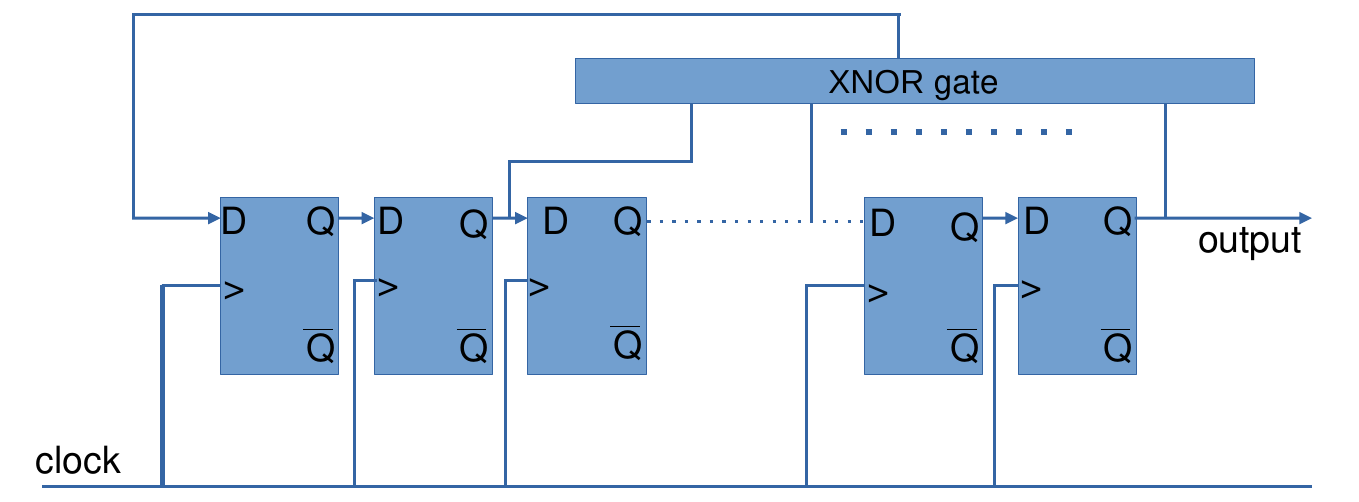}
    \caption{Technique of generating LRNG using cascaded flip flops and XNOR.}
    \label{fig:lfsr}
\end{figure}

LFSR, a type of sequential digital circuit, utilizes clock-driven flip-flops in its operation. It is composed of a collection of D flip-flops, and the quantity of these flip-flops determines the number of bits in the LFSR's seed. Specifically, a $d$-bit LFSR requires $d$ number of D flip-flops. If appropriate taps (XORing position of bits in the initial seed) are chosen, LFSR produces a sequence of bits that are random with a repeatability cycle of ($2^d-1$) \cite{Peter1996Efficient}. The operational concept of the LRNG is illustrated in the FIG. \ref{fig:lfsr}.

Inspired by hardware implementation, one can also simulate these LFSR on a computer using primitive polynomials. There are two types of LFSR based on the type of XORing operation: Galois field-based and Fibonacci numbers-based \cite{LFSR2023}. Galois field based have an easier software implementation, while Fibonacci based has an easier hardware implementation. This is primarily because Fibonacci based LFSRs don't require additional `AND' operations and, thus, more easily weave into the fabric of FPGA.

One of the major disadvantages of the LRNG is that they are linear. Hence, algorithms to catch such correlations exist in the literature. One such algorithm is Berlekamp-Massey Algorithm (BMA) which is the basis of the Linear Complexity Test (LCT) in NIST test suites. Since BMA targets to decipher the generation process of LFSR rather than the output periodicity ($2^d-1$), it can decipher a $d$-bit LFSR after measuring just $2d$ bits with a very high probability. Hence, we add non-linearity via XOR operation to explore the nuances between LCT and the quality of the LFSR bit stream.

\subsection{Testing random numbers using NIST statistical test suite} \label{sec:NIST}
To check the quality of LRNG, we treat the NIST test suite as a gold standard for checking randomness. It is a sufficient criterion, as suggested in the literature. The test suite is a hypothesis testing suite where $H_0$ is null hypothesis that sequence is random while $H_{a}$ is alternate hypothesis that sequence is not random. In this suite, we majorly focus on Type-I and Type-II errors as depicted below in the TABLE \ref{table:NIST} and highlighted in \cite{NISTGitHub}.
\begin{table}[h]
    \centering
    \begin{tabular}{|c|c|c|}
    \hline
    True situation & Accept $H_{0}$ & Accept $H_{a}$ \\ \hline
    $H_{0}$ is true & No error & Type-I  \\ \hline
    $H_{a}$ is true & Type-II & No error  \\ \hline
    \end{tabular}
    \caption{Test static $p$ value is equivalent to $\alpha$ where $\alpha$ is a threshold chosen for Type I errors.}
    \label{table:NIST}
\end{table} 

One can assign a threshold $\alpha$ and $\beta$ for Type-I and Type-II errors, respectively. The test static $p$ value defines the $\alpha$ value, and $\alpha$ and $\beta$ are connected. Thus, keeping a check on one is enough, and for NIST, it is $\alpha$. Lower $\alpha$ values means higher $\beta$ values. 

The test suite checks for certain statistical correlations. A couple of tests in the suite check for disparity between zeroes and ones in a binary sequence for short range and long range. A few of the tests are a variant of auto-correlation tests. Others check for patterns using some pre-decided templates or based on Fourier transforms. The most stringest amongst these tests is LCT as it targets the process of LFSR rather than the output. Other than LCT, LFSRs pass all other 14 statistical tests pretty easily. Also, to avoid misrepresenting these tests using different input parameters, NIST recommends specifying these parameters upfront. In our case, we have used default input parameters, and the test-statistic $p$ is set to the default 0.01.

\subsection{Numerical approach}
Before implementing on hardware, we generated a variety of Fibonacci number based LRNG on software using primitive polynomials. The first variation is on different length strings, whether a 4-bit or 8-bit LFSR. The code used to generate the LRNG is obtained from the GitHub repository \cite{nikesh_bajaj}. To generate a random bit stream in our case, we have augmented the Matlab code, which generates the desired number of bits ($d$) for some defined random input seeds ($s$). For ease of understanding of further variations tested, we use the convention $\mathcal{L}(d,s)$ where $d$ is width of seed and $s$ is initial seed value. In the numerical test bench, we have generated 55M bit stream to check against NIST (pre-requisite of 55$\times$1 M, as recommended in the manual).

For the second variation, we utilize two randomly generated bit sequences derived from $\mathcal{L}(d,s_1)$ and $\mathcal{L}(d,s_2)$. The non linearity to the bit stream is introduced via XORing the two outputs. In the third case, we generalise it with inputs XOR($\mathcal{L}(d_1,s_1)$, $\mathcal{L}(d_2,s_2)$). The premise of choosing $d$ values is that the numbers should be co-prime. We have discussed two sub-cases in this category. The first variant involves using consecutive $d$ values, and the second variant involves using consecutive prime $d$ values. 

\subsection{Previous attempts}
Previous attempts in the realm of hardware implementation of random numbers \cite{wang2020100} have explored different ideas to overcome the limitations of algorithm-based randomness. Two veins branch out in this direction. One generating random numbers using noise (chaos based random bits). This noise could be because of clock jitter \cite{demir2020comparative}, \cite{majzoobi2011fpga} circuit noise \cite{santoro2009fly}, \cite{xu2016high} etc. The other generating LFSR based random numbers by introducing sufficient non-linearities\cite{zode2019fpga}. This work follows the second vein and introduces a minimum non linearity such that it provides a low computational cost on the FPGA for QKD applications. This minimum non-linearity can be theoretically backed easily with similar approaches to \cite{fuster2014weak}.

\section{Results and Discussion} \label{sec:result}
The following subsections show results for all four variants in tabular form. Broadly speaking, the NIST test suite has 15 statistical tests, and LCT has the stringiest requirement against the LFSR generated bit streams and hence is explicitly mentioned in another column. We have generated the bit stream first on software, and later, hardware implementation is performed only for cases where it passes the NIST test suite for all 15 statistical tests. We have mentioned all the less-sensitive tests (14 of 15) under the column name NIST-tests, while LCT is explicitly mentioned in another column in the tabular representation.

\subsection{Numerically generated RNG test results} \label{sec:NISTResult}
The numerical simulation was performed using primitive polynomials. The results of the simulation are shown in TABLE \ref{table:LFSRInitialSeed}, \ref{table:SameXOR}, \ref{table:ConsXOR}, and \ref{table:PrimeXOR}.

\subsubsection*{Variant 1: $\mathcal{L}(d,s)$}
This is the simplest case. Here, initialized shift register $\mathcal{L}(d,s)$ with the seed $s$ and taps which generate the maximum length bit-sequence ($2^d-1$) and tested for quality of randomness. We observed that only in case of $d=128$, 14 out of 15 NIST tests are passed. However, the LCT tests still fails. For smaller $d$ values, all the 15 tests failed. The results are shown in TABLE ~\ref{table:LFSRInitialSeed}.
\begin{table}[h]
    \centering
    \begin{tabular}{|c|c|c|}
    \hline
    $d$-bit & NIST-tests & LCT \\ \hline
    8 & Failed & Failed \\ \hline
    16 & Failed & Failed \\ \hline
    24 & Failed & Failed \\ \hline
    32 & Failed & Failed \\ \hline
    64 & Failed & Failed \\ \hline
    128 & Passed & Failed \\ \hline
    \end{tabular}
    \caption{Test result for LFSR $\mathcal{L}(d,s)$}
    \label{table:LFSRInitialSeed}
\end{table}

Results for the case of $\mathcal{L}(d,s)$ bolsters our faith in the quality of the randomness test suite. The failure for $d=128$ is evident as the bit stream is generated via the linear operations. This holds true for any $d$ or $s$ value. Hence, we employ a single XOR operator as a non-linear function to enhance the quality of randomness required for LCT.

\subsubsection*{Variant 2: XOR($\mathcal{L}(d,s_1)$, $\mathcal{L}(d,s_2)$)}
In the second variation, we generate two random-bit sequences using two different seeds ($s$ value) but the same $d$-bit LFSR. This also doesn't improve the quality of randomness, as verified by the test suite. A few reasons contribute to this failure. Starting with XOR being a linear operation between two same $d$ LFSRs. Further, BMA (disguised as LCT in NIST tests) targets the process rather than the bit stream. Finally, LFSR is generated with primitive polynomials implying both BMA and LFSR are seed-agnostic. Hence, the results are similar to what has been discussed in the previous case. The results for same $d$ values are shown in TABLE ~\ref{table:SameXOR}.
\begin{table}[h]
    \centering
    \begin{tabular}{|c|c|c|}
    \hline
    $d$ & NIST-tests & LCT \\ \hline
    8 & Failed & Failed \\ \hline
    16 & Failed & Failed \\ \hline
    24 & Failed & Failed \\ \hline
    32 & Failed & Failed \\ \hline
    64 & Failed & Failed \\ \hline
    128 & Passed & Failed \\ \hline
    \end{tabular}
    \caption{Test result for XOR($\mathcal{L}(d,s_1)$, $\mathcal{L}(d,s_2)$).}
    \label{table:SameXOR}
\end{table}

\subsubsection*{Variant 3.1: XOR($\mathcal{L}(d_1,s_1)$, $\mathcal{L}(d_2,s_2)$)}
In the third category, we generalise the above XOR operation to consecutive $d's$, ($d_1$ and $d_2$ being consecutive are co-prime). The sequence passes all the quality checks, except LCT for the XOR of $d=24$ and $d=25$. However, the results for $d=128$ and $d=129$ are different. It is based on the fact that the XOR operation introduces enough non-linearity such that LCT isn't able to catch it. Even the other NIST tests pass for as low as $d_1=24$ and $d_2=25$ as shown in TABLE ~\ref{table:ConsXOR}.
\begin{table}[h]
    \centering
    \begin{tabular}{|c|c|c|c|}
    \hline
    $d_1$ & $d_2$ & NIST-tests & LCT \\ \hline
    8 & 9 & Failed & Failed \\ \hline
    16 & 17 & Failed & Failed \\ \hline
    24 & 25 & Passed & Failed \\ \hline
    32 & 33 & Passed & Failed \\ \hline
    64 & 65 & Passed & Failed \\ \hline
    128 & 129 & Passed & Passed \\ \hline
    \end{tabular}
    \caption{Results for Variant 3.1: XOR($\mathcal{L}(d_1,s_1)$, $\mathcal{L}(d_2,s_2)$)} 
    \label{table:ConsXOR}
\end{table}

\subsubsection*{Variant 3.2: XOR($\mathcal{L}(d_1,s_1)$, $\mathcal{L}(d_2,s_2)$)}
In an attempt to find lower $d$ values for hardware implementation, we looked at another variant of XOR($\mathcal{L}(d_1,s_1)$, $\mathcal{L}(d_2,s_2)$), with $d_1$ and $d_2$ to be consecutive prime numbers. However, this didn't provide any significant advantage compared to Variant 1 with results in TABLE ~\ref{table:ConsXOR} for our QKD application. Also, the results did improve compared to the previous variant. Specifically, for values of $d_{1}=7$ and $d_{2}=11$, the generated bit sequences pass all NIST tests except for the LCT, which initially showed failures. For XOR($\mathcal{L}(127,s_1)$, $\mathcal{L}(131,s_2)$) passed all the tests and thus, can be used for suitable application. The results are shown in TABLE  \ref{table:PrimeXOR}.
\begin{table}[h]
    \centering
    \begin{tabular}{|c|c|c|c|}
    \hline
    $d_1$-bit & $d_2$-bit & NIST-tests & LCT \\ \hline
    3 & 5 & Failed & Failed \\ \hline
    5 & 7 & Failed & Failed \\ \hline
    7 & 11 & Passed & Failed \\ \hline
    11 & 13 & Passed & Failed \\ \hline
    ... & ... & ... & ... \\ \hline
    113 & 127 & Passed & Failed \\ \hline
    127 & 131 & Passed & Passed \\ \hline
    \end{tabular}
    \caption{Results for Variant 3.2: XOR($\mathcal{L}(d_1,s_1)$, $\mathcal{L}(d_2,s_2)$)}
    \label{table:PrimeXOR}
\end{table}

As evident from the preceding four variations of LRNGs, it is clear that XOR($\mathcal{L}(128,s_1)$, $\mathcal{L}(129,s_2)$) and XOR($\mathcal{L}(127,s_1)$, $\mathcal{L}(131,s_2)$) passes all the 15 tests (14 NIST-tests + LCT). To further investigate, we took 100 sets of random sequences for these two cases. We found that XOR($\mathcal{L}(128,s_1)$, $\mathcal{L}(129,s_2)$) fails 1.53\% of the individual tests, while XOR($\mathcal{L}(127,s_1)$,  $\mathcal{L}(131,s_2)$) fails 1.40\% of any individual tests. Passing these tests is a proof that our test statistics $p$ value set to 0.01 is correct. Moreover, for both the variants, the interval of acceptance proportions given the significance level $\alpha$ and number of tested sequences $k$ \cite{marton2015interpretation} lies within
\begin{equation*}
    (1-\alpha) \pm 3\times \sqrt{\frac{{\alpha}(1-{\alpha})}{k}}.
\end{equation*}

For the above case, $\alpha=0.01$ and $k$=100 and lies within the range [0.99$\pm$0.0099499]. With all the consistency checks and good results, we implement both XOR($\mathcal{L}(128,s_1)$, $\mathcal{L}(129,s_2)$) and XOR($\mathcal{L}(127,s_1)$, $\mathcal{L}(131,s_2)$) on FPGA/hardware for our QKD source generation.

\subsection{Hardware implementation and test results}\label{sec:HardwareImplementation}
Utilizing hardware like FPGA is a necessity for QKD experiments, primarily because of the generation of voltage pulses to control/trigger the operation of any circuit/hardware. For this purpose, we prefer Arty 7 FPGA Evaluation Kit, Artix 7 35T FPGA \cite{xlinix2015Arty7} from AMD to implement the proposed LRNG method. The design of this board is specifically focused on providing a remarkably flexible Micro Blaze Soft Processing System. The board incorporates an on-chip analog-to-digital converter and provides an ample number of I/O options for generating and acquiring voltage signals. Consequently, there is no need for an additional daughter board to facilitate interfacing with real-time signals. In addition to its specifications, this board boasts a compact form factor, affordability, and suitability for our operations. Additionally, the board is a low cost (\$159) and provides 100 MHz clock rate.

We have chosen Very high speed integrated circuit hardware Description Language (VHDL) codes for our implementation. One can use Verilog or another high level language like C or Python for programming FPGA. The generated random voltage pulses are time tagged and subsequently generated bits are tested against NIST test suite. The resulting outcomes are presented in the TABLE \ref{table:FPGA128129} and \ref{table:FPGA127131}. The VHDL code implemented on FPGA for the current article can be obtained from the Ref. \cite{fpgacode}.

\subsubsection{Test setup}
\begin{figure}[h]
    \centering
    \includegraphics[width=8cm]{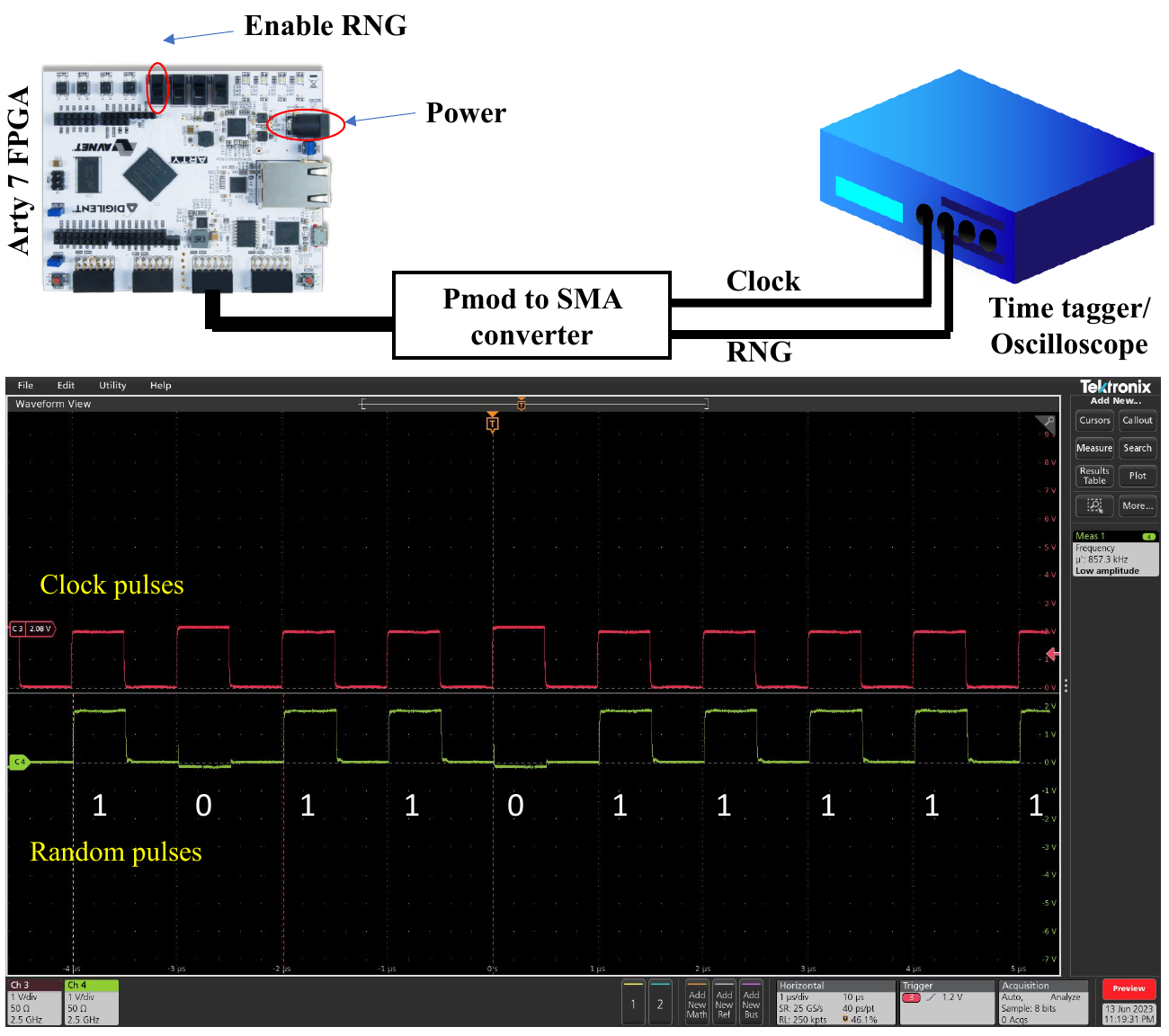}
    \caption{Timing waveform of random voltage pulses and clock pulses generated from the FPGA board and obtained using an oscilloscope. When both the and clock pulses are active simultaneously, it indicates a bit value `1'. Conversely, when only the clock pulse is present without any accompanying random pulse, it represents a bit  value `0'.}
    \label{fig:FPGA}
\end{figure}

The VHDL code for generating random numbers via shift register is fed to Arty 7 FPGA board. The setup for generating time stamps and corresponding voltage pulses is shown in FIG. \ref{fig:FPGA}. Along with random voltage pulses, we also generate clock pulses for referencing to assign random pulses as bit 0 and 1. In this setup, the FPGA board's output is derived from its I/O ports, featuring Pmod connectors. Typically to generate random bits, the clock and random voltage signal produced by the FPGA board must be fed into the time tagger module. However, since the time tagger module is equipped with an SMA connector, the use of a Pmod to SMA converter  shown in FIG. \ref{fig:FPGA} becomes imperative to establish the necessary connection. Furthermore, in QKD setup, the laser diode driver circuit also incorporates an SMA connector. Consequently, the utilization of a Pmod to SMA converter is also crucial to seamlessly integrate the FPGA board's signals with the laser diode driver circuit.

The timing waveform in FIG. \ref{fig:FPGA} illustrates the waveform of the random voltage pulses and reference clock pulses. We utilize a time tagger (IDQuantique ID900) to accurately track the timestamps of the clock and random pulses. This enables precise recording and synchronization of their timing information.

\subsubsection{NIST results for FPGA implemented random bit stream} 
To test the random pulses generated from the FPGA board for XOR($\mathcal{L}(128,s_1)$, $\mathcal{L}(129,s_2)$) and XOR($\mathcal{L}(127,s_1)$, $\mathcal{L}(131,s_2)$), we connect both the ports (clock and random signals) to the IDQuantique ID900 time-tagger. It records the time of the arrival of these pulses. The time stamps recorded by the time-tagger are post-processed in Matlab to obtain a raw binary bit stream.

In the first step, we record time stamps of random pulses and clock pulses for XOR($\mathcal{L}(128,s_1)$), $\mathcal{L}(129,s_2)$). The clock rate for random voltage generation is set to 1 MHz for hardware implementation. The overall data consists of (55 sets$\times$ 1 M) random bits, where each set is recorded for 1 second and is concatenated. The results obtained are in good agreement with the numerically obtained data. We have used the ``randperm()'' function, which is built on the Ziggurat algorithm variant incorporating MT-19937 as a generator. The TABLE  \ref{table:FPGA128129} showcases the results from different shuffling referred as Random-1/2/3/4/5 while other concatenations are obvious. The last row entry under the column ``None" means that we generated a random bit stream of 55 M length in one go. In all of these cases, all 15 NIST tests successfully passed. Similarly for XOR($\mathcal{L}(127,s_1)$, $\mathcal{L}(131,s_2)$), the corresponding results are depicted in the TABLE ~\ref{table:FPGA127131}. 
\begin{table}[h]
    \centering
    \begin{tabular}{|c|c|c|c|}
    \hline
    Exposure time & Concatenation & NIST-tests & LCT \\ \hline
    1 s & Serial & Passed & Passed \\ \hline
    1 s & Reverse & Passed & Passed \\ \hline
    1 s & Random-1 & Passed & Passed \\ \hline
    1 s & Random-2 & Passed & Passed \\ \hline
    1 s & Random-3 & Passed & Passed \\ \hline
    1 s & Random-4 & Passed & Passed \\ \hline
    1 s & Random-5 & Passed & Passed \\ \hline
    55 s & None & Passed & Passed \\ \hline
    \end{tabular}
    \caption{Test results for different shufflings of  XOR($\mathcal{L}(128,s_1)$, $\mathcal{L}(129,s_2)$), obtained from Arty 7}
    \label{table:FPGA128129}
\end{table}

\begin{table}[h]
    \centering
    \begin{tabular}{|c|c|c|c|}
    \hline
    Exposure time & Concatenation & NIST-tests & LCT \\ \hline
    1 s & Serial & Passed & Passed \\ \hline
    1 s & Reverse & Passed & Passed \\ \hline
    1 s & Random-1 & Passed & Passed \\ \hline
    1 s & Random-2 & Passed & Passed \\ \hline
    1 s & Random-3 & Passed & Passed \\ \hline
    1 s & Random-4 & Passed & Passed \\ \hline
    1 s & Random-5 & Passed & Passed \\ \hline
    55 s & None & Passed & Passed \\ \hline
    \end{tabular}
    \caption{Test results for different shufflings of XOR($\mathcal{L}(127,s_1)$, $\mathcal{L}(131,s_2)$) obtained from Arty 7}
    \label{table:FPGA127131}
\end{table}

Throughout the hardware implementation, the selection of an optimal clock rate proved to be a crucial parameter due to the limitations in synthesizing certain code configurations on the FPGA board. We systematically generated random numbers for 1 MHz clock rate using FPGA. The device consumes 64 mW of power for XOR ($\mathcal{L}(128,s_1)$, $\mathcal{L}(129,s_2)$) and 63 mW for XOR($\mathcal{L}(127,s_1)$, $\mathcal{L}(131,s_2)$). 

\subsection{Application of FPGA based RNG in QKD}\label{sec:QKD source}
The major application for LFSR in QKD applications comes into the picture for P\&M protocols. Our application is mainly dedicated to the BB84 protocol implementation which is one of the well-studied P\&M and popular QKD protocols. The operational functioning behind P\&M based QKD protocols is to use a particular property of photons (say, polarisation/phase) and use it to generate quantum states in mutually unbiased bases, send them to another party who measures them, and form a key. Say, Alice chooses a polarisation degree of freedom to encode and share the key with Bob for communication. Alice needs to prepare her states which are indistinguishable (arbitrary) in four different polarisation quantum states, namely horizontal $\ket{H}$, vertical $\ket{V}$, diagonal $\ket{D}$, anti-diagonal$\ket{A}$. To prepare this signal, Alice uses four different laser diodes (one for each quantum state) and combines them. However, while combining, we need to make sure that laser circuits for all four laser diodes should fire randomly so that the signal is unbiased to all four quantum states. This requirement is a pre-requisite for unconditional security of any P\&M-based QKD protocol.

\begin{figure}[h]
    \centering
    \includegraphics[width=8.5cm]{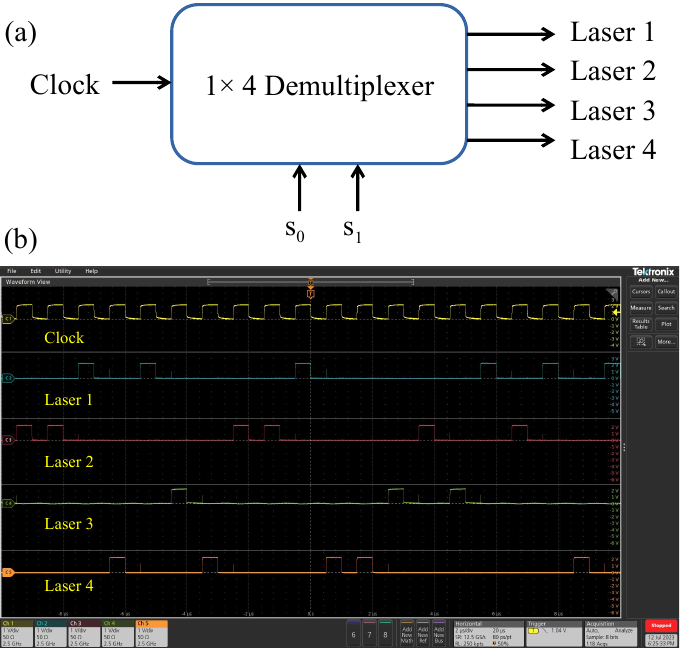}
    \caption{a) Schematic of generating four TTL pulses using 1$\times$4 demultiplexer using the random bits generated from LFSR as a select line $s_0$ and $s_1$. The select line enables one laser at a time and performs the demultiplexing. b) Four color representing four random TTL pulses generated from FPGA to drive four corresponding laser diodes, each diodes corresponds to polarisation states ($\ket{H}$, $\ket{V}$, $\ket{D}$, $\ket{A}$).}
    \label{fig:QKD}
\end{figure}

\begin{figure}[h]
    \centering
    \includegraphics[width=8.5cm]{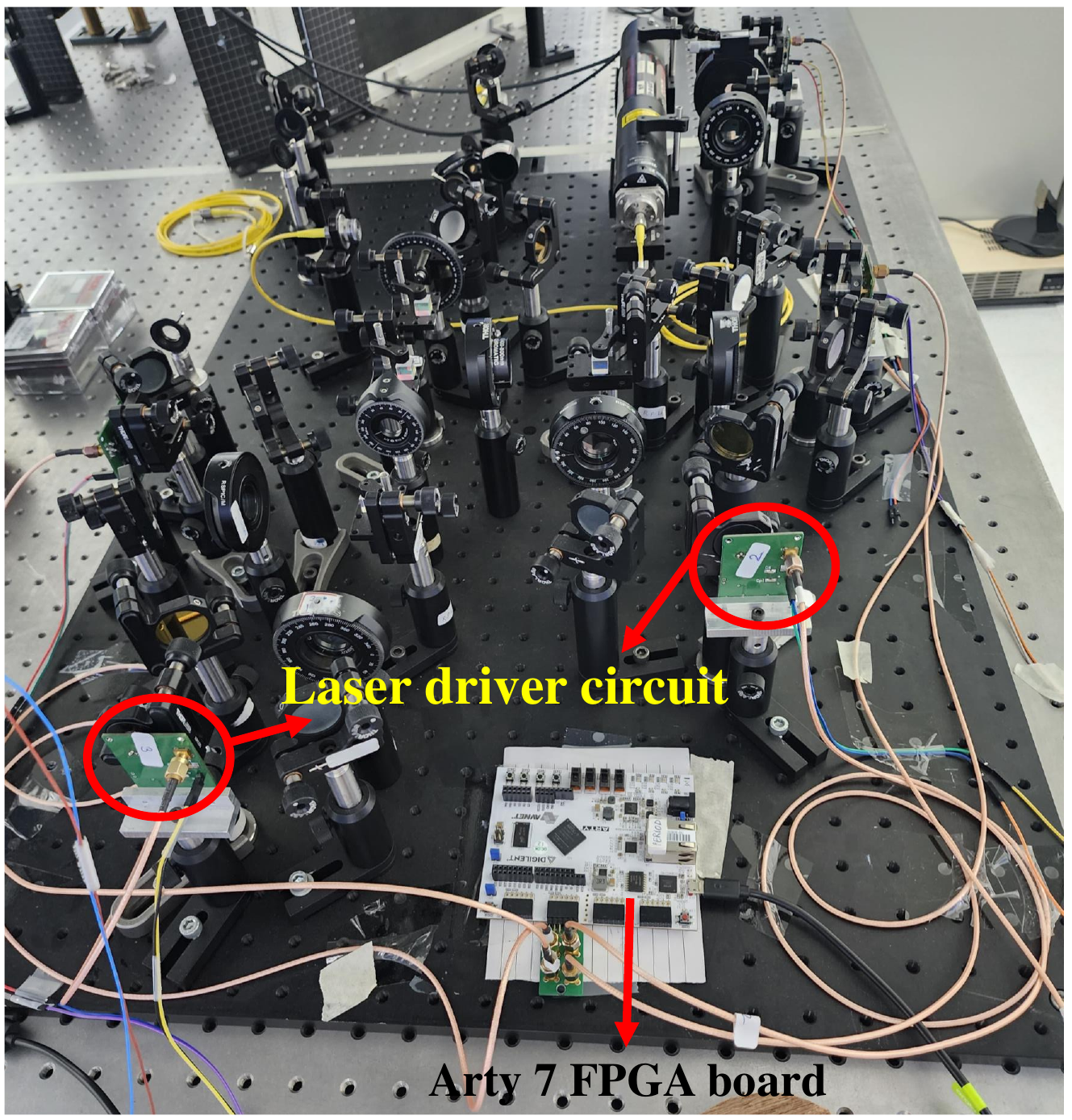}
    \caption{The optical setup for BB84 QKD setup which includes a Arty 7 FPGA board to generate RNG, which drives the laser diode driver circuit for QKD experiments.}
    \label{fig:optical}
\end{figure}

To meet the specified requirements, we have developed a laser driver circuit that utilizes random voltage pulses generated from an FPGA. Additionally, we have incorporated a 1$\times$4 demultiplexer in the FPGA, which is controlled by randomly generated select lines derived from LFSR generated random bits. This demultiplexer ensures that only one output port of the FPGA is enabled at any given time, thereby allowing for the activation of a single laser diode at a time. By implementing this demultiplexing scheme, we can achieve the generation of four random polarization states in FIG. \ref{fig:QKD}. The schematic diagram illustrating the demultiplexing scheme is presented in FIG. \ref{fig:QKD} (top). The random voltage pulses generated from Arty 7 for driving the laser diode driver circuit is shown in FIG. \ref{fig:QKD} (bottom). The optical setup of the QKD transmitter is shown in FIG. \ref{fig:optical}. Here, we have used laser diode driver circuit (marked in red circle), integrated with other optical components of BB84 setup \cite{Ayan2021Experimental}. Information bits are encoded in polarisation and then sent to Bob via quantum channel.

\section{Outlook}\label{sec:outlook}
In this work, we have studied the application of LFSR based RNG implemented on FPGAs to be used in QKD applications. Within the QKD framework, these RNG can also be used to provide robustness against electromagnetic interference while transferring data to a trusted node via Serial Advanced Technology Attachment (SATA) protocol \cite{guha2016design}. Outside of QKD, it could be used for carrying out vulnerability analysis of FPGA based computer vision applications \cite{chadjiminas2015field}. On the premise that there are no specific guidelines on generating a robust non-linear Boolean function, a natural extension of this work could be to look in this direction. LFSRs on FPGAs are being tested for Internet of Things (IoT) applications using Artificial Neural Networks \cite{teodoro2021fpga}. In this work, we improved the quality of LRNG by modifying the generation process of RNG. The proposed method and proposed $d$ for LFSRs are, one can say, are sufficient to pass the NIST statistical test suite; however, they are not the limit. Depending upon the application requirement and availability of the FPGA board, there are several possible improvements in this technique. Some improvements that are foreseen are listed below:
\begin{enumerate}
    \item One can generate a new seed everytime the device boots such that the linear correlations cannot be deciphered if one models the data for too long.
    \item Another open direction is to use higher $d$ values, not restricting up to 131. This could be used for other applications. Say, for example, one can choose XOR($\mathcal{L}(785,s_1)$, $\mathcal{L}(786,s_2)$) which passes not only the default block length parameter (m=500) in LCT but passes LCT upto m=3100. This direction of inquiry is strictly application and resource dependent.
    \item In the current work we have synthesised VHDL code of LRNG in FPGA for different clock rate(1 MHz, 5MHz, 10 MHz, 15 MHz, 20 MHz, and 25 MHz) for 2 cases XOR($\mathcal{L}(128,s_1)$, $\mathcal{L}(129,s_2)$) and XOR($\mathcal{L}(127,s_1)$, $\mathcal{L}(131,s_2)$). However, hardware implementation and verification on FPGA was done only for 1 MHz as per our current requirement for QKD application. As the code is synthesizable for higher bit-rates, this technique promises compatibility for setups with higher key rate. Depending on hardware constraints and use case, and one can generate the bit stream in GHz range which is suitable for video-conferencing using QKD applications.
    \item Increasing complexity in a controlled manner is another way to circumvent going to higher d values to pass LCT test for higher m values. The amount of controlled complexity that can be introduced is dependent on both hardware and theoretically proof of non-linearity. For example operations like XOR(XOR($\mathcal{L}(d_1,s_1)$, $\mathcal{L}(d_2,s_2)$), XOR($\mathcal{L}(d_3,s_3)$, $\mathcal{L}(d_4,s_4)$)) using four different $d$'s and $s$'s.  
    \item On the quality of randomness testing front, one can check for chi-square hypothesis testing against other statistical test suites, say, Dieharder, TestU01, ENT. One could also use other alternatives for hypothesis testing, like the Kolmogorov-Smirnoff test or Borel normality, depending on how stringent the randomness is for specific applications. 
\end{enumerate} 

\section{Conclusion}\label{sec:conclusion}
In this article, we proposed a method to generate LRNG for QKD applications which passes all th NIST statistical test suites and can be easily theoretically backed. The number of tests failed also cross-verifies the meaning of test specific $p$-value. We verified various cases of single XOR operation, which can introduce enough non-linearity to pass the LCT test. We have shown that amongst the variants XOR($\mathcal{L}(128,s_1)$, $\mathcal{L}(129,s_2)$) and XOR($\mathcal{L}(127,s_3)$, $\mathcal{L}(131,s_4)$) passed the test for 100 sets with better accuracy and hence, is a better random number generator. Compared to the consensus on the quality of the bit stream, our random bit stream passes stringent conditions (15/15 tests). Further, the proposed method was implemented on a low-cost FPGA board and verified that those random pulses, too, passed the test. On the application side, we use our method for the weak-coherent pulses-based source to study the BB84 protocols.

\section*{Acknowledgement}
The authors like to acknowledge the funding support from the Department of Science and Technology (DST), India through QuEST program. The authors also acknowledge initial discussions with Dr. Ayan Biswas, York Centre for Quantum Technologies, School of Physics Engineering and Technology, University of York. 

\section*{Data Availability}
The data that support the findings of this study are available from the corresponding author upon reasonable request.

\section*{Disclosures}
The authors declare no conflicts of interest related to this article.

\bibliography{main}

\end{document}